\documentclass[conference]{IEEEtran}
\usepackage{amsmath}
\usepackage[pdftex]{graphicx}
\usepackage{graphicx}
\usepackage{pstricks}
\usepackage{amsfonts}
\usepackage{tabularx}
\newtheorem{Lemma}{Lemma}

\usepackage{amssymb}
\usepackage{multirow}
\usepackage{epsfig}
\usepackage{epstopdf}

\usepackage[noadjust]{cite} 

\begin{document}
\title{Secrecy Fairness Aware NOMA for Untrusted Users}
\author{\IEEEauthorblockN{Sapna Thapar$^{1}$, Deepak Mishra$^{2}$, and Ravikant Saini$^{1}$}
\IEEEauthorblockA{$^{1}$Department of Electrical Engineering, Indian Institute of Technology Jammu, India\\
$^{2}$Department of Electrical Engineering (ISY), Link\"oping University, Sweden\\
Emails: 2018ree0019@iitjammu.ac.in, deepak.mishra@liu.se, ravikant.saini@iitjammu.ac.in
}}
\maketitle

\begin{abstract}
Spectrally-efficient secure non-orthogonal multiple access (NOMA) has recently attained a substantial research interest for fifth generation development. This work explores crucial security issue in NOMA which is stemmed from utilizing the decoding concept of successive interference cancellation. Considering untrusted users, we design a novel secure NOMA transmission protocol to maximize secrecy fairness among users. A new decoding order for two users' NOMA is proposed that provides positive secrecy rate to both users. Observing the objective of maximizing secrecy fairness between users under given power budget constraint, the problem is formulated as minimizing the maximum  secrecy outage probability (SOP) between users. In particular, closed-form expressions of SOP for both users are derived to analyze secrecy performance. SOP minimization problems are solved using pseudoconvexity concept, and optimized power allocation (PA) for each user is obtained. Asymptotic expressions of SOPs, and optimal PAs minimizing these approximations are obtained to get deeper insights. Further, globally-optimized power control solution from secrecy fairness perspective is obtained at a low computational complexity and, asymptotic approximation is obtained to gain analytical insights. Numerical results validate the correctness of analysis, and present insights on optimal solutions. Finally, we present insights on global-optimal PA by which fairness is ensured and gains of about $55.12\%$, $69.30\%$, and $19.11\%$, respectively are achieved, compared to fixed PA and individual users' optimal PAs.
\end{abstract}
\section{Introduction}
Non-orthogonal multiple access (NOMA) is envisaged as a potential breakthrough for fifth generation (5G) networks because of the possibility of serving multiple users within same resource block \cite{nomasurvey}. Conversly, the broadcast nature of wireless communication at transmitter, and decoding concept of successive interference cancellation (SIC) at receiver, in NOMA, causes security challenge on the information-carrying signal. Therefore, the research on security issues in NOMA networks has attained great attention among 5G researchers. The integration of NOMA and physical layer security (PLS) has been observed as a new research frontier towards providing spectrally-efficient and secure wireless communication \cite{liu2017enhancing}. Still, despite merits, the design process includes security challenge of wiretapping in the presence of untrusted users.
\subsection{Related Art}
Motivated by the spectral efficiency improvement by NOMA, \cite{nomasurvey} has addressed research contributions in power-domain NOMA. Stimulated by potential of PLS, \cite{pls} has summarized existing research works on PLS techniques. Recently, innumerable researchers have concentrated on PLS in NOMA. PLS in large-scale networks has been studied in \cite{liu2017enhancing} where a protected zone around base station (BS) is designed to retain an eavesdropper-free region. Secure NOMA with multiple users against eavesdropper has been discussed in \cite{zhang2016secrecy} for single-input single-output network. A joint beamforming scheme has been introduced in \cite{ding2017spectral}, where confidential data is transmitted to intended user only. Secrecy of a cooperative NOMA system with a decode-and-forward and an amplify-and-forward relay against eavesdropper has been analyzed in \cite{8245831}. A secrecy beamforming scheme that exploits artificial noise (AN) to enhance secrecy of NOMA in the presence of eavesdropper has been presented in \cite{lv2018secure}. Besides eavesdroppers, NOMA itself has inherent security issue which is caused due to SIC based decoding at receiver. Regarding this, recently \cite{basepaper} has considered a system where near user is assumed to be trusted, whereas far user as untrusted, and investigated secrecy of only trusted user against untrusted node.

\begin{figure}[!t]\label{system_model}
\centering
\includegraphics[scale=.4]{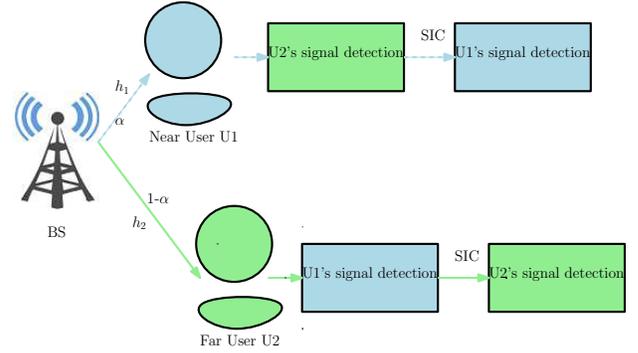}
\caption{Illustration of downlink NOMA system with two untrusted users where decoding order is changed for far user  compared to conventional approach.}
\end{figure}

\subsection{Research Gap and Motivation}
As noted, existing works have considered different PLS techniques such as AN aided strategy \cite{liu2017enhancing}, \cite{lv2018secure}, optimal power allocation (PA) \cite{zhang2016secrecy}, beamforming \cite{ding2017spectral}, and cooperative relaying \cite{8245831} to improve secrecy of NOMA against external eavesdroppers. Taking conventional decoding order concept in NOMA into account, two key steps are followed: (1) observing near user signal (with better channel conditions) as noise, far user (with poorer channel conditions) decodes its own signal first, and after decoding, it may apply SIC and decode signal of near user \cite{basepaper}; (2) near user first decodes signal associated to far user, applies SIC, and then decodes its own signal. \textit{As an outcome, near and far users, respectively, have access of far and near user which is a critical security concern in NOMA implementation between untrusted users.} Considering this issue, \cite{basepaper} has assumed only far user as untrusted and analyzed the secrecy performance of trusted (near) user.

As inferred, the system, assuming all users as untrusted is more challenging and practical scenario for designing a secure network. Untrusted users' model is a more hostile situation, where all users do not have mutual trust and each user focuses on achieving secure communication from BS in the presence of untrusted users \cite{saini2016ofdma}, \cite{saini2019subcarrier}, \cite{7987792}. \textit{Towards this end, we investigate secure NOMA protocol from positive secrecy rates standpoint for both untrusted users, which to the best of our knowledge, has been an open problem in literature.} 

\subsection{Key Contributions}
Considering a system with one BS and two untrusted users, \textit{a novel secure NOMA protocol is designed to maximize secrecy fairness between untrusted users}. Main contributions are as follows: (1) A new decoding order for two users' NOMA is proposed that provides positive secrecy rate to both users. (2) Analytical expressions of secrecy outage probability (SOP) for both users, and their asymptotic approximations have been derived. (3) Closed-form expressions of optimal PA minimizing SOP, have been obtained using pseudoconvexity of SOP at both users. (4) Global-optimal PA solution to a problem minimizing the maximum SOP between users under given power budget constraint, is obtained. (5) Numerical results validate analytical derivations, present insights on optimal solutions, and analyze performance gains with proposed model.

\section{Proposed Secure NOMA Protocol}
\subsection{System Model}
We consider downlink NOMA system where BS communicates with two untrusted users (Fig. $1$). Our two users consideration will shed light on the proof of concept, however, the protocol can be extended to a system with multiple users. Each node is equipped with single antenna \cite{basepaper}. We denote U$1$ and U$2$ as near and far user, respectively. $h_{i}$ is denoted as Rayleigh fading channel gain coefficient from BS to U$i$ where $i\in \{1,2\}$. All the channels from BS to users are assumed to be independent, and follow small scale fading accompanied with path loss effects, such that channel power gain $|h_{i}|^{2}$ experience exponential distribution with mean $\lambda_{i}=L_{c}d_{i}^{-n}$ where $L_{c}$, $n$, and $d_{i}$, respectively, denote path loss constant, path loss exponent and distance between BS and U$i$. Assuming statistical channel state information is known at BS, U$1$ and U$2$, respectively, are considered as strong and weak users. Channel power gains are sorted as $|h_{1}|^{2}>|h_{2}|^{2}$. A fixed amount of transmit power $P_{t}$ is allocated from BS to users. $\alpha$ denotes the PA coefficient, i.e., the fraction of $P_{t}$ allocated to U$1$. Remaining fraction, i.e, $(1-\alpha)$ is allocated to U$2$. 

Applying power-domain NOMA principle, BS broadcasts superposition of information signals $x_{1}$ and $x_{2}$ of U$1$ and U$2$, respectively, and then the transmitted signal is $\sqrt{\alpha P_{t}} x_{1}$+ $\sqrt{(1-\alpha)P_{t}} x_{2}$ \cite{basepaper}. The received signals $y_{1}$ and $y_{2}$, respectively, at U$1$ and U$2$, from BS are given as \cite{basepaper}
\begin{equation}
y_{1} = h_{1}( \sqrt{\alpha P_{t}}x_{1}+ \sqrt{(1-\alpha)P_{t}} x_{2})  + n_{1},
\end{equation}
\begin{equation}
y_{2} = h_{2}( \sqrt{\alpha P_{t}}x_{1}+ \sqrt{(1-\alpha)P_{t}} x_{2})  + n_{2},
\end{equation} 
where $n_{1}$ and $ n_{2}$ denote additive white Gaussian noise (AWGN) with mean $0$ and variance $\sigma^{2}$ at both users. We assume ideal SIC based decoding at receivers where interference from other user is perfectly cancelled at legitimate user. However, in real scenarios, perfect SIC cannot be readily satisfied due to various implementation problems such as error propagation. Therefore, imperfect SIC model where residual interference from imperfectly decoded user exists after SIC, is highly realistic to explore secure NOMA which has been considered in the extended version of this work \cite{gc2019noma}.   
\subsection{Proposed Decoding Order for Untrusted NOMA}
In secure NOMA protocol, signal of U$2$ must be protected from U$1$, and vice-versa. Before discussing secure protocol, we first present insights on how conventional decoding order is inefficient for providing secrecy at both users in untrusted scenario. Considering conventional NOMA (Section I(B)), the received signal-to-interference-plus-noise-ratio $\Gamma_{ij}$ at U$i$ when signal of U$i$ is decoded by U$j$ (for $i\in \{1,2\}, j \in \{1,2\}$) is given as \cite{basepaper}
\begin{align}
\Gamma_{21} = \frac{(1-\alpha)| h_{1}|^{2}}{\alpha|h_{1}|^{2} + \frac{1}{\rho_{t}} }, \quad
\Gamma_{22} = \frac{(1-\alpha)|h_{2} |^{2}}{\alpha|h_{2}|^{2} + \frac{1}{\rho_{t}} }, \nonumber\\
\Gamma_{11} = \alpha\rho_{t}| h_{1}|^{2},\quad 
\Gamma_{12} = \alpha\rho_{t}|h_{2}| ^{2},
\end{align}
where $\rho_{t}\stackrel{\Delta}{=}P_{t}$/$\sigma^{2}$ is BS transmit signal-to-noise ratio (SNR). Secrecy rates $R_{s1}$ and $R_{s2}$ for U$1$ and U$2$ can be given by
\begin{equation}\label{Rs1}
R_{s1} = R_{11} - R_{12}, \quad R_{s2} = R_{22} - R_{21}, 
\end{equation}
where $R_{11}$, $R_{12}$, $R_{21}$ and $R_{22}$, respectively, are given as \cite{tse2005fundamentals}
\begin{align} \label{R11R12}
R_{11} = \log_{2}(1+\Gamma_{11}), \quad \nonumber
R_{12} = \log_{2}(1+\Gamma_{12}),\\
R_{22} =\log_{2}(1+\Gamma_{22}), \quad
R_{21} =\log_{2}(1+\Gamma_{21}).
\end{align}

The condition $R_{11}> R_{12}$ required for positive secrecy rate at U$1$, simplified as $\Gamma_{11} > \Gamma_{12}$ gives a feasible condition $|h_{1}|^{2}>| h_{2}|^{2}$. This ensures positive secrecy rate at U$1$. Next, for positive $R_{s2}$ at U$2$, $R_{22} > R_{21}$, simplified as $\Gamma_{22}>\Gamma_{21}$ results an infeasible condition $|h_{2}|^{2} >| h_{1}|^{2}$ because channel power gains are assumed as $|h_{1}|^{2} >| h_{2}|^{2}$. Thus, positive secrecy rate is not achieved at U$2$. Hence, the conventional decoding order cannot be considered for untrusted NOMA. 

Now, with the goal of providing positive secrecy rate to both users,  \textit{we propose a new decoding order, according to which both U$1$ and U$2$ first decode signal associated to other user, and then decode its own signal after performing SIC.} Compared to the conventional NOMA, the decoding order is changed for the far user only. As a result, $\Gamma_{ij}$ are
\begin{align}
\Gamma_{21} = \frac{(1-\alpha)|h_{1}|^{2}}{\alpha|h_{1}|^{2}+\frac{1}{\rho_{t}}}, \quad
\Gamma_{12} = \frac{\alpha|h_{2}|^{2}}{(1-\alpha)|h_{2}|^{2} + \frac{1}{\rho_{t}}},  \nonumber \\
\Gamma_{11} = \alpha\rho_{t} |h_{1}|^{2}, \quad
\Gamma_{22} = (1-\alpha)\rho_{t}|h_{2}|^{2}.
\end{align}

For positive $R_{s1}$, $R_{11} > R_{12}$, simplified as $\Gamma_{11} > \Gamma_{12}$ gives 
\begin{equation}\label{alpha1}
\alpha < 1 + \frac{|h_{1}|^{2}-|h_{2}|^{2}}{|h_{1}|^{2}|h_{2}|^{2}\rho_{t}}.
\end{equation}
 
Thus, positive secrecy rate can be ensured at U$1$. Similarly, for positive $R_{s2}$, $R_{22}> R_{21}$, simplified as $\Gamma_{22}>\Gamma_{21}$ gives 
\begin{equation}\label{alpha2}
\alpha >  \frac{|h_{1}|^{2}-|h_{2}|^{2}}{|h_{1}|^{2}|h_{2}|^{2}\rho_{t}}.\\
\end{equation}

Observing \eqref{alpha1} and \eqref{alpha2}, it can be concluded that proposed decoding order is efficient to provide positive secrecy rate to both users in untrusted NOMA, provided $\frac{|h_{1}|^{2}-|h_{2}|^{2}}{|h_{1}|^{2}|h_{2}|^{2}\rho_{t}} < \alpha< 1$. 
\section{Secrecy Performance Analysis}
Next we derive analytical expressions of SOP and investigate  optimal PAs minimizing SOPs for both U$1$ and U$2$. 
\subsection{Exact Secrecy Outage Probability}
The SOP is defined as the probability that maximum achievable secrecy rate at each user falls below a target secrecy rate. Denoting  $s_{oi}$ as SOP for U$i$, now we derive SOPs analytically. 
\subsubsection{Near user} Considering achievable and target secrecy rate of U$1$ as $R_{s1}$ \eqref{Rs1} and $R_{s1}^{th}$, respectively, $s_{o1}$ is given as
\begin{align}\label{sop_near_user}
s_{o1} &= \text{Pr}\{R_{s1} <  R_{s1}^{th}\}=\text{Pr}\Big\{\frac{1+\Gamma_{11}}{1+\Gamma_{12}} < \Pi_{1}\Big\},\nonumber \\
&=\text{Pr}\Big\{|h_{1}|^{2} < \frac{\Pi_{1}|h_{2}|^{2}}{(1-\alpha)\rho_{t}|h_{2}|^{2} + 1} + A_{1}\Big\},\nonumber \\
&= \int_{0}^{\infty} F_{\mid h_{1} \mid ^{2}}\bigg(\frac{\Pi_{1}| h_{2}|^{2}}{(1-\alpha)\rho_{t} |h_{2}|^{2}+1}+A_{1}\bigg)f_{| h_{2} |^{2}}(y_{1}) dy_{1},\nonumber \\
&=\!1-\! \frac{1}{\lambda_{2}} \int_{0}^{\infty} \!\exp\bigg\{\!\frac{-\Pi_{1}y_{1}}{((1-\alpha)\rho_{t} y_{1}+1)\lambda_{1}}\! - \!\frac{y_{1}}{\lambda_{2}} - \frac{A_{1}}{\lambda_{1}} \bigg\}dy_{1},
\end{align}
where $\text{Pr}\{.\}$ is  denoted for the probability measure, $\Pi_{1} \stackrel{\Delta}{=} 2^{R_{s1}^{th}}$, $A_{1} \stackrel{\Delta}{=} \frac{\Pi_{1}-1}{\alpha\rho_{t}}$, $F_{|h_{1}|^{2}}(x)$ and $f_{\mid h_{2} \mid ^{2}}(x)$ are the cumulative distribution function (CDF) and probability density function (PDF), respectively, of exponentially distributed random channel power gain $|h_{1}|^{2}$ and $|h_{2}|^{2}$, respectively.
\subsubsection{Far user} 
Considering $R_{s2}$ \eqref{Rs1} and $R_{s2}^{th}$ as achievable and target secrecy rate, respectively, of U$2$, $s_{o2}$ is stated as
\begin{equation}\label{sop_far_user}
\begin{split}
s_{o2}  &= \text{Pr}\{R_{s2} <  R_{s2}^{th}\} = \text{Pr}\Big\{\!\log_{2}\Big(\frac{1+\Gamma_{22}}{1+\Gamma_{21}}\Big) < R_{s2}^{th}\!\Big\},\\
&=\text{Pr}\Big\{\frac{1+\Gamma_{22}}{1+\Gamma_{21}} < \Pi_{2}\Big\} \!=\! \text{Pr}\Big\{\!|h_{2}|^{2} \!<\! \frac{\Pi_{2}| h_{1} | ^{2}}{\alpha\rho_{t}| h_{1} |^{2} + 1} \!+\! A_{2}\!\Big\},\\
&= \int_{0}^{\infty} F_{\mid h_{2} \mid ^{2}}\bigg(\frac{\Pi_{2}|h_{1}|^{2}}{\alpha\rho_{t}|h_{1} |^{2}+1}+A_{2}\bigg)f_{| h_{1} |^{2}}(y_{2}) dy_{2},\\
&=\!1\!-\!\frac{1}{\lambda_{1}}\int_{0}^{\infty} \exp\bigg\{\frac{-\Pi_{2}y_{2}}{(\alpha\rho_{t} y_{2}+1)\lambda_{2}} - \frac{y_{2}}{\lambda_{1}}-\frac{A_{2}}{\lambda_{2}} \bigg\} dy_{2},
\end{split}
\end{equation}
where $\Pi_{2} \triangleq 2^{R_{s2}^{th}}$, $A_{2} \triangleq \frac{\Pi_{2}-1}{(1-\alpha)\rho_{t}}$, $F_{|h_{2}|^{2}}(x)$ and $f_{\mid h_{1} \mid ^{2}}(x)$ are the CDF and PDF of $|h_{2}|^{2}$ and $|h_{1}|^{2}$, respectively.
\subsection{Secrecy Outage Probability Minimization}
\subsubsection{Near User} The SOP minimization problem for U$1$, considering $s_{o1}$ \eqref{sop_near_user} as a function of $\alpha$, can be stated as 
\begin{align}\label{problem_form_nearuser}
(J1): \underset{\alpha}{\text{minimize }} s_{o1}, \text{ subject to } (C1): 0<\alpha<1.
\end{align}
The optimality of problem $(J1)$ is asserted by Lemma $1$. 
\begin{Lemma}
$s_{o1}$ is pseudoconvex function of $\alpha$. 
\end{Lemma}
\begin{IEEEproof}
Denoting integrand of $s_{o1}$ \eqref{sop_near_user}, as $I_{1}$, we obtain
\begin{equation}
I_{1} = \frac{1}{\lambda_{2}}\exp\bigg\{- \frac{\Pi_{1}y_{1}}{((1-\alpha)\rho_{t} y_{1}+1)\lambda_{1}} - \frac{y_{1}}{\lambda_{2}}-\frac{(\Pi_{1}-1)}{\alpha \rho_{t}\lambda_{1}}\bigg\}.
\end{equation}
The second-order derivative of $\log(I_{1})$ with respect to $\alpha$ is 
\begin{equation}
\begin{split}
 \frac{d^{2}\log(I_{1})}{d\alpha^{2}} = - \Big(\frac{2 (\Pi_{1}-1)}{\rho_{t}\lambda_{1}\alpha^{3}} +
 \frac{2 \Pi_{1}\rho_{t}^{2} y_{1}^{3}}{\lambda_{1}((1-\alpha)\rho_{t} y_{1} + 1)^{3}}\Big),
\end{split}
\end{equation}
which is decreasing and shows $I_{1}$ is a logarithmically concave function. Because log-concavity is preserved under integration \cite{boyd2004convex}, the integral function in \eqref{sop_near_user} is also log-concave function. Considering pseudoconcave property \cite[Lemma 5]{mishra2016joint} of log-concave function, the integral function of \eqref{sop_near_user} is pseudoconcave. The negative of pseudoconcave function is pseudoconvex function \cite{bazaara1979nonlinear}. Hence, $s_{o1}$ is pseudoconvex function of $\alpha$.
\end{IEEEproof} 

We apply golden section search (GSS) algorithm \cite{chang2009n} to find optimal solution $\alpha_{1}^{*}$ which minimizes $s_{o1}$. GSS algorithm considers pseudoconvex function $s_{o1}$, lower and upper bounds of $\alpha$, i.e., $\alpha_{lb}$ and $\alpha_{ub}$, respectively, as input. It provides optimized solution $\alpha_{1}^{*}$ and corresponding minimized $s_{o1}$ as outputs. Firstly, $\alpha_{lb}=0$ and $\alpha_{ub}=1$ are considered and algorithm searches along $\alpha$ with $\epsilon<<1$, where $\epsilon$ is acceptable tolerance. Algorithm functions by a reduction in search interval with a fixed ratio of $0.618$ at the end of each iteration. Algorithm terminates when the search length is less than a pre-determined tolerance level \cite{chang2009n}. 
\subsubsection{Far User}
$s_{o2}$ minimization problem can be stated as 
\begin{align}\label{problem_form_faruser}
(J2):\underset{\alpha}{\text{minimize }} s_{o2}, \text{  subject to } (C1).
\end{align}
The feasibility of unique solution is proved in Lemma $2$.
\begin{Lemma}
\textit{$s_{o2}$ is a pseudoconvex function of $\alpha$.} 
\end{Lemma}
\begin{IEEEproof}
Denoting integrand of $s_{o2}$ \eqref{sop_far_user}, as $I_{2}$, we obtain 
\begin{equation}
I_{2} = \frac{1}{\lambda_{1}}\exp\bigg\{- \frac{\Pi_{2}y_{2}}{(\alpha\rho_{t} y_{2}+1)\lambda_{2}} - \frac{y_{2}}{\lambda_{1}}-\frac{(\Pi_{2}-1)}{(1-\alpha) \rho_{t}\lambda_{2}}\bigg\}.
\end{equation}

Observing $\frac{d^{2}\log(I_{2})}{d\alpha^{2}}$ is monotonically decreasing, $I_{2}$ is also a log-concave function, and similar to the proof in Lemma $1$, $s_{o2}$ is also a pseudoconvex function of $\alpha$. 
\end{IEEEproof}

The optimal solution $\alpha_{2}^{*}$ of $(J2)$ is evaluated using GSS algorithm by considering $s_{o2}$ function as input. 
\subsection{Asymptotic Approximations: SOP and Optimization}
In aforementioned analysis, SOP minimization problems have been solved numerically due to the complexity of derived expressions. 
Next, we present asymptotic approximations of SOPs and optimal PAs to gain analytical insights.
\subsubsection{Near User}
Asymptotic expression of $s_{o1}$, i.e., $\hat s_{o1}$, can be obtained by setting $((1-\alpha)\rho_{t} y_{1}+1)$ $\approx$ $(1-\alpha)\rho_{t} y_{1}$ for $\rho_{t}\gg1$ in \eqref{sop_near_user}. Accordingly, $\hat s_{o1}$ can be given as
\begin{align} \label{sop1_asy}
\hat s_{o1} &\!=\!1\!-\!\exp\bigg\{ \frac{- \Pi_{1}}{(1-\alpha)\rho_{t}\lambda_{1}} - \!\frac{(\Pi_{1}-1)}{\alpha\rho_{t}\lambda_{1}} \!\bigg\}\! \int_{0}^{\infty}\!\frac{\exp\{-\frac{y_{1}}{\lambda_{2}}\}}{\lambda_{2}}dy_{1}, \nonumber \\
&= 1 - \exp\bigg\{\frac{\Pi_{1}+\alpha-1} {\alpha(\alpha-1)\rho_{t}\lambda_{1}}\bigg\}.
\end{align}

The $\hat s_{o1}$ minimization problem can be formulated as
\begin{equation}
\begin{aligned}
(J3): \underset{\alpha}{\text{minimize}}
\quad  \hat s_{o1}, 
\quad \text{subject to} \quad (C1).\\
\end{aligned}
\end{equation}

Lemma $3$ gives the optimal solution for $(J3)$. 
\begin{Lemma}
\textit{The asymptotic optimal PA $\hat \alpha_{1}$, that minimizes $\hat s_{o1}$, can be given as
\begin{equation}\label{optimal_a_asy_sop1}
\hat \alpha_{1} = - (\Pi_{1}-1) + \sqrt{(\Pi_{1}(\Pi_{1} -1)}.
\end{equation}}
\end{Lemma}
\begin{IEEEproof}
Since $\hat \alpha_{1}$ is obtained by minimizing $\hat s_{o1}$ \eqref{sop1_asy}, second-order derivative of $\hat s_{o1}$ with respect to $\alpha$ is given as
\begin{align}
\frac{\mathrm{d}^2\hat s_{o1}}{\mathrm{d}\alpha^2}= &\Big\{\dfrac{2\Pi_{1}}{\lambda_1 \rho_{t}\left(1-\alpha\right)^3} -\dfrac{2\left(1-\Pi_{1}\right)}{\lambda_1 \rho_{t}\alpha^3} \nonumber\\ 
&-\Big(-\dfrac{1-\Pi_{1}}{\lambda_1 \rho_{t} \alpha^2}-\dfrac{\Pi_{1}}{\lambda_1 \rho_{t} \left(1-\alpha\right)^2}\Big)^2\Big\} \nonumber\\
& \times \exp\Big\{\frac{1-\Pi_{1}}{\lambda_1\rho_{t}\alpha}-\frac{\Pi_{1}}{\lambda_1 \rho_{t}\left(1-\alpha\right)}\Big\}, 
\end{align}
which does not imply monotonic behaviour. We set $\frac{\mathrm{d}\hat s_{o1}}{\mathrm{d}\alpha}=0$, and obtain $\hat \alpha_{1} = - (\Pi_{1}-1) \pm \sqrt{(\Pi_{1}(\Pi_{1} -1)}$. Note that $\hat \alpha_{1}  = - (\Pi_{1}-1) - \sqrt{(\Pi_{1}(\Pi_{1} -1)}$ is negative, and therefore, infeasible. Hence, $\hat \alpha_{1}$ minimizing $\hat s_{o1}$ is given as \eqref{optimal_a_asy_sop1}.
\end{IEEEproof}
\subsubsection{Far User}
Asymptotic approximation of $s_{o2}$, i.e., $\hat s_{o2}$ obtained using  $(\alpha\rho_{t} y_{2}+1)$ $\approx$ $\alpha\rho_{t} y_{2}$ in \eqref{sop_far_user} for high $\rho_{t}$, is 
\begin{align}\label{sop2_asy}
\hat s_{o2} &=1-\exp\bigg\{ \frac{- \Pi_{2}}{\alpha\rho_{t}\lambda_{2}} - \frac{(\Pi_{2}-1)}{(1-\alpha)\rho_{t}\lambda_{2}} \bigg\} \int_{0}^{\infty}\frac{\exp\{-\frac{y_{2}}{\lambda_{1}}\}}{\lambda_{1}}dy_{2}, \nonumber \\
&=1 - \exp\bigg\{\frac{\Pi_{2}-\alpha} {\alpha(\alpha-1)\rho_{t}\lambda_{2}}\bigg\}.
\end{align}

Now $\hat s_{o2}$ minimization problem for U$2$ can be stated as 
\begin{equation}
\begin{aligned}
(J4):\underset{\alpha}{\text{minimize}}
& &  \hat s_{o2}, 
& & \text{subject to} \quad (C1).\\
\end{aligned}
\end{equation}

Optimal solution for minimizing $\hat s_{o2}$ is given by Lemma $4$.
\begin{Lemma}
\textit{The asymptotic optimal PA $\hat \alpha_{2}$ minimizing $\hat s_{o2}$ can be given as
\begin{equation}\label{optimal_a_asy_sop2}
\hat \alpha_{2} =  \Pi_{2} - \sqrt{(\Pi_{2}(\Pi_{2} -1)}.
\end{equation}}
\end{Lemma}
\begin{IEEEproof}
By setting, $\frac{\mathrm{d}\hat s_{o2}}{\mathrm{d}\alpha} = 0$, $\hat \alpha_{2} =  \Pi_{2} \pm \sqrt{(\Pi_{2}(\Pi_{2} -1)}$ is obtained. Here $\hat \alpha_{2} = \Pi_{2} + \sqrt{(\Pi_{2}(\Pi_{2} -1)}$ is infeasible, because it forces $R_{s2}^{th}<0$ which is infeasible since secrecy rate cannot be negative. Hence, $\hat \alpha_{2}$ for U$2$ is given as \eqref{optimal_a_asy_sop2}.
\end{IEEEproof} 
\section{Maximization of Secrecy Fairness }
Next we formulate secrecy fairness optimization problem and, investigate globally-optimized PA to maximize fairness. 
\subsection{Optimization Formulation}
Using \eqref{sop_near_user} and \eqref{sop_far_user}, the secrecy fairness maximization problem which minimizes the maximum SOP between users under BS transmit power budget constraint can be stated as 
\begin{align}\label{problem_formulation}
(J5): \underset{\alpha}{\text{minimize}} \quad \max[s_{o1},s_{o2}],\quad \text{subject to} \quad (C1). 
\end{align}
Using $x_{c} \stackrel{\Delta}{=} \max[ s_{o1},s_{o2}]$, $(J5)$ is formulated equivalently as 
\begin{align}\label{problem_formulation1}
& (J6): \underset{\alpha,x_{c}}{\text{minimize }} x_{c}, \quad \text{subject to} \quad (C1), \nonumber \\ 
&  (C2): s_{o1}\leq x_{c}, \quad (C3): s_{o2}\leq x_{c},
\end{align}
where $(C2)$ and $(C3)$ comes from the definition of max$[\cdot]$.
\subsection{Power Control for Optimizing $\min$-$\max$ Secrecy Outage}
Since $(J6)$ is nonconvex problem because of $(C2)$ and $(C3)$ nonconvex constraints, we solve it by analyzing optimal candidates that are characterized by Karush-Kuhn-Tucker (KKT) conditions \cite{ravindran2006engineering}. Global-optimal PA is given by Lemma $5$.
\begin{Lemma}
\textit{The global-optimal solution $\alpha_{sop}^{*}$ of $(J6)$, which minimizes the maximum SOP between users, is given as
\begin{equation}\label{optimal_a_asy_sop}
\!\alpha_{sop}^{*} \! \stackrel{\Delta}{=}\!\underset{\alpha \in \{\alpha_{1}^{*}, \alpha_{2}^{*}, \alpha_{3}^{*}\}}{\mathrm{argmin}}\! \max[s_{o1},s_{o2}],
\end{equation}
where $\alpha_{1}^{*}, \alpha_{2}^{*}, \alpha_{3}^{*}$ are obtained using GSS by minimizing $s_{o1}, s_{o2}$ ( Section III(B)), and solving $s_{o1}=s_{o2}$, respectively.}
\end{Lemma}
\begin{IEEEproof}
We consider boundary constraint $(C1)$ implicit and associate Lagrange multipliers $\eta_{1}$ with $(C2)$ and $\eta_{2}$ with $(C3)$. Hence, Lagrangian function $\mathcal{L}$ of $(J6)$ can be given as
\begin{equation}\label{lagragian_function1}
\mathcal{L} \stackrel{\Delta}{=}  x_{c} + \eta_{1}[ s_{o1} - x_{c}] + \eta_{2}[ s_{o2} -  x_{c}].
\end{equation}

The corresponding KKT conditions are given by constraints $(C1)$, $(C2)$ and $(C3)$. The dual feasibility conditions are given as $\eta_{1}\geq 0$ and $\eta_{2}\geq 0$. The subgradient conditions are obtained as
$\frac{d\mathcal{L}}{d x_{c}}  =  1-\eta_{1}-\eta_{2}= 0, \quad
\frac{d\mathcal{L}}{d\alpha}  =  \eta_{1} \frac{d s_{o1}}{d\alpha}  + \eta_{2}\frac{d s_{o2}}{d\alpha} = 0$. The complementary slackness conditions are given as
\begin{equation}\label{sop1=sop2_1}
 \eta_{1} [ s_{o1} - x_{c}] = 0, \quad
 \eta_{2} [ s_{o2} -  x_{c}] = 0.
\end{equation}

Here exists three cases. \emph{Case 1: $\eta_{1} > 0$ and $\eta_{2}= 0$}, implies $\frac{\mathrm{d} s_{o1}}{\mathrm{d}\alpha}=0$ which results same solution of $s_{o1}$ minimization \eqref{problem_form_nearuser}, i.e., $\alpha=\alpha_{1}^{*}$. \emph{Case 2: $\eta_{2} > 0$ and $\eta_{1}= 0$}, implies $\frac{\mathrm{d}s_{o2}}{\mathrm{d}\alpha}=0$, which results same solution of $s_{o2}$ minimization \eqref{problem_form_faruser} of U$2$, i.e., $\alpha = \alpha_{2}^{*}$. \emph{Case 3: $\eta_{1} > 0$ and $\eta_{2} > 0$}, implies $ s_{o1} = s_{o2}$ \eqref{sop1=sop2_1}, which shows equal SOP for both users, and gives $\alpha = \alpha_{3}^{*}$. Thus, $(J6)$ has three candidates, i.e., $\alpha_{1}^{*}$ and $\alpha_{2}^{*}$ for minimizing $s_{o1}$ and $s_{o2}$, respectively, and $\alpha_{3}^{*}$ is obtained from $s_{o1}=s_{o2}$ condition. As a result, global-optimal $\alpha_{sop}^{*}$ to $(J6)$ problem is obtained at the optimal candidate for which maximum SOP between users is minimum.
\end{IEEEproof}
\subsection{Closed-form Approximation of Optimal Power Allocation}
In above analysis, the min-max SOP optimization problem has been solved numerically. To gain analytical insights, next the asymptotic closed-form approximation of global-optimal PA for high SNR is derived. Here, the asymptotic secrecy fairness maximization problem can be formulated as
\begin{align}
(J7): \underset{\alpha}{\text{minimize }} \max[\hat s_{o1}, \hat s_{o2}], \quad \text{subject to }\quad (C1),
\end{align}
Considering $\hat x_{c} \stackrel{\Delta}{=} \max[\hat s_{o1},\hat s_{o2}]$, $(J7)$ can be rewritten as 
\begin{equation}
\begin{aligned}\label{prob_form_asym}
(J8): \quad \underset{\alpha, \hat x_{c}}{\text{minimize}} \quad \hat x_{c}, \quad
\text{subject to} \quad (C1),\\
(C4): \hat s_{o1}\leq \hat x_{c}, \quad
(C5): \hat s_{o2}\leq \hat x_{c},
\end{aligned}
\end{equation}
where $(C4)$ and $(C5)$ also comes from definition of max$[\cdot]$. Globally optimized solution of $(J8)$ is given in Lemma \ref{lemma_2}.
\begin{Lemma} \label{lemma_2}
\textit{Asymptotic global-optimal PA $\hat \alpha_{sop}$ of min-max problem $(J8)$ that maximizes secrecy fairness is given by
\begin{equation}\label{optimal_asy_a}
\hat \alpha_{sop} \stackrel{\Delta}{=} \underset{\alpha \in \{\hat \alpha_{1}, \hat \alpha_{2}, \hat \alpha_{3}\}}{\mathrm{argmin}}\max[\hat s_{o1},\hat s_{o2}],
\end{equation}
where $\hat \alpha_{1}, \hat \alpha_{2}$, are obtained by minimizing $\hat s_{o1}, \hat s_{o2}$ (Section III(C)), respectively, and $\hat \alpha_{3}$ is derived by solving $\hat s_{o1}=\hat s_{o2}$.}
\end{Lemma}
\begin{IEEEproof}
Associating lagrange multipliers $\mu_{1}$ and $\mu_{2}$, respectively, with $(C4)$ and $(C5)$, the Lagrangian function $\mathcal{\hat L}$ can be written as
\begin{equation}\label{lagragian_function}
\mathcal{\hat L} \stackrel{\Delta}{=} \hat x_{c} + \mu_{1}[\hat s_{o1} - \hat x_{c}] + \mu_{2}[\hat s_{o2} - \hat x_{c}].
\end{equation}

The corresponding KKT conditions are obtained by constraints $(C4)$ and $(C5)$. The dual feasibility conditions are given as $\mu_{1}\geq 0$ and $\mu_{2}\geq 0$ using \eqref{prob_form_asym} and \eqref{lagragian_function}. The subgradient conditions are given as $
\frac{\mathrm{d}\mathcal{\hat L}}{\mathrm{d}\hat x_{c}}  =  1-\mu_{1}-\mu_{2}= 0, \quad
\frac{\mathrm{d}\mathcal{\hat L}}{\mathrm{d}\alpha}  =  \mu_{1} \frac{\mathrm{d}\hat s_{o1}}{\mathrm{d}\alpha}  + \mu_{2}\frac{\mathrm{d}\hat s_{o2}}{\mathrm{d}\alpha} = 0$. The complementary slackness conditions are given as
\begin{equation}\label{sop1=sop2}
 \mu_{1} [\hat s_{o1} - \hat x_{c}] = 0, \quad
 \mu_{2} [\hat s_{o2} - \hat x_{c}] = 0.
\end{equation}

Similar to the numerical proof (Section IV(B)), here also three cases exist by analyzing KKT conditions. \emph{Case 1: $\mu_{1} > 0$ and $\mu_{2}= 0$}, implies $\frac{\mathrm{d}\hat s_{o1}}{\mathrm{d}\alpha}=0$ and results $\alpha = \hat \alpha_{1}$ \eqref{optimal_a_asy_sop1} from $\hat s_{o1}$ minimization. \emph{Case 2: $\mu_{2} > 0$ and $\mu_{1}= 0$}, implies $\frac{\mathrm{d}\hat s_{o2}}{\mathrm{d}\alpha}=0$. This gives $\alpha = \hat \alpha_{2}$ \eqref{optimal_a_asy_sop2} as from $\hat s_{o2}$ minimization. \emph{Case 3: $\mu_{1} > 0$ and $\mu_{2} > 0$}, implies $\hat s_{o1} = \hat s_{o2}$ from \eqref{sop1=sop2} and it gives $\hat \alpha_{3}$ which is obtained as
\begin{equation}
\hat \alpha_{3} = \frac{ \Pi_{2}\lambda_{1} + \lambda_{2} (1-\Pi_{1})}{\lambda_{1} + \lambda_{2}}.
\end{equation}
Since three candidates exist for minimization problem $(J8)$, the global-optimal solution $\hat \alpha_{sop}$  is obtained at the candidate for which maximum SOP between users is minimum.
\end{IEEEproof}
\section{Numerical Investigations}
For generating numerical results, downlink of NOMA system with a BS and two users is considered. Near and far user distances from BS are considered as $d_{1}=50$ meter and $d_{2}=100$ meter, respectively. Noise signal for both users follows Gaussian distribution with a noise power of $-60$ dBm. Small scale fading follows exponential distribution with $1$ mean value \cite{basepaper}. $L_{c}=1$ and $n=2.5$ are taken. The simulation results are sampled over $10^6$ randomly generated channel realizations utilizing Rayleigh distribution for both the users. For GSS algorithm, $\epsilon=0.01$. $\rho_{r}$ is assumed as received SNR in decibels (dB) at U$2$. SOP is considered as performance metric to evaluate system performance.
\subsection{Validation of Analysis}
\begin{figure}[!t]
\centering
\includegraphics[scale=.43]{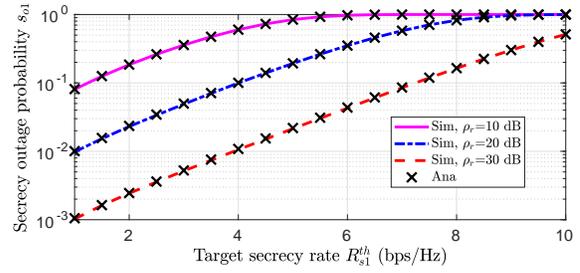}
\caption{Validation of U$1$'s secrecy outage probability $s_{o1}$ analysis.}
\label{validation_00}
\vspace{-0.7mm}
\end{figure}

We first validate the closed-form expressions of SOP derived in section III. Fig. \ref{validation_00} presents validation results of $s_{o1}$ with $R_{s1}^{th}$ for various $\rho_{r}$. $\alpha=0.5$ is taken. A close match between analytical and simulation results confirms the accuracy of analysis of $s_{o1}$ with a RMSE of the order of $10^{-4}$. We observe from results that increasing $R_{s1}^{th}$  increase $s_{o1}$. Considering the definition that the outage happens when the users' maximum achievable secrecy rate falls below a target rate, it is obvious that increasing target secrecy rates at user increases SOP. Also, we observe that increasing $\rho_{r}$ decreases $s_{o1}$. This is because the achievable secrecy rates at users increase by increasing SNR, and hence, for a fixed target secrecy rate, SOP decreases. 
\subsection{Impact of variation of far user distance}
\begin{figure}[!t]
\centering
\includegraphics[scale=.43]{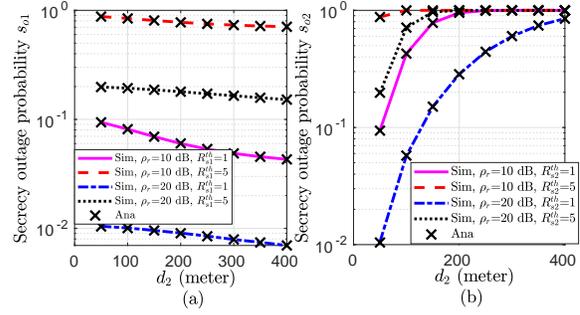}
\caption{Variation of SOP versus U$2$'s distance $d_{2}$ (a) $s_{o1}$, and (b) $s_{o2}$.}
\label{distance_variation}
\end{figure}
\begin{figure}[!t]
\centering
\includegraphics[scale=.43]{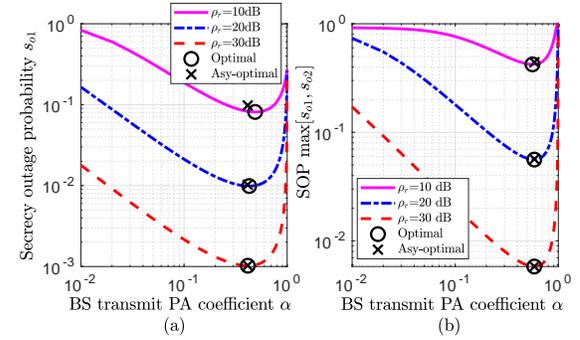}
\caption{(a) Optimal $s_{o1}$ and $\alpha$ analysis for U$1$ at $R_{s1}^{th}=1$ and, (b) Optimal secrecy fairness and $\alpha$ analysis at $R_{s1}^{th}=1$, $R_{s2}^{th}=1$.}
\label{optimal_outage}
\vspace{-0.8mm}
\end{figure}
Fixing $d_{1}=50$ meter, the impact of variation of $d_{2}$ from BS on achievable SOP is presented in Fig. \ref{distance_variation}. Fig. \ref{distance_variation}(a), demonstrate the effect of increasing $d_{2}$ on $s_{o1}$, show that $s_{o1}$ decreases with the increase in $d_{2}$. The is because, increasing $d_{2}$ implies a decrease in achievable data rate at U$2$ which results an improvement in secrecy rate at U$1$, and hence, SOP at U$1$ decreases. Also, decrease in data rate at U$2$ implies decrease in secrecy rate at U$2$ which increases SOP for U$2$ as shown in Fig. \ref{distance_variation}(b). It is noted that increasing the distance from BS to U$2$ has an contradicting effect on $s_{o1}$ and $s_{o2}$. Hence, we conclude that achievable SOP depends on distances of users.
\subsection{Optimal Design Insights}
Now optimal SOP is investigated in Fig. \ref{optimal_outage}(a) and Fig. \ref{optimal_outage}(b), which validate pseudoconvex nature of $s_{o1}$ and $\max[s_{o1}, s_{o2}]$, respectively, with $\alpha$. The numerical optimal PAs are obtained using GSS algorithm. Asymptotic analysis is also verified with numerical results at high SNR, i.e., $\rho_{r} \geq 20$ dB. Here we observe that $\alpha$ decides PA to users, which highly effects SOPs. Hence, for given system parameters, the appropriate PA to users can ensure optimal secure communication system. 
\begin{figure}[!t]
\centering
\includegraphics[scale=.43]{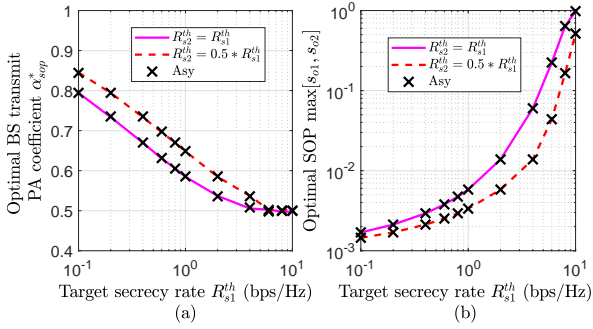}
\caption{(a) Global-optimal PA $\alpha_{sop}^{*}$ with U$1$'s target secrecy rate $R_{s1}^{th}$, and (b) optimal secrecy fairness analysis with $R_{s1}^{th}$ at $\rho_{r}=30$ dB.}
\label{a_comp}
\end{figure}
Next, global-optimal $\alpha_{sop}^{*}$ that provides secrecy fairness between users is shown in Fig. \ref{a_comp}(a) as a function of $R_{s1}^{th}$ for various $R_{s2}^{th}$. Results indicate that there exist one and only one optimal $\alpha$ for each target secrecy rate pair ($R_{s1}^{th}$, $R_{s2}^{th}$). We observe that increasing $R_{s1}^{th}$, $\alpha_{sop}^{*}$ decreases, whereas the optimal SOP obtained from min-max optimization problem increases as shown in Fig. \ref{a_comp}(b). It is also noted that lower value of $R_{s2}^{th}$ compared to $R_{s1}^{th}$ provides an improvement in SOP. Hence, we conclude that $\alpha_{sop}^{*}$ that provides secrecy fairness to users highly depends on target secrecy rate pair $(R_{s1}^{th}, R_{s2}^{th})$. 
\subsection{Performance Comparison}
\begin{figure}[!t]
\centering
\includegraphics[scale=.43]{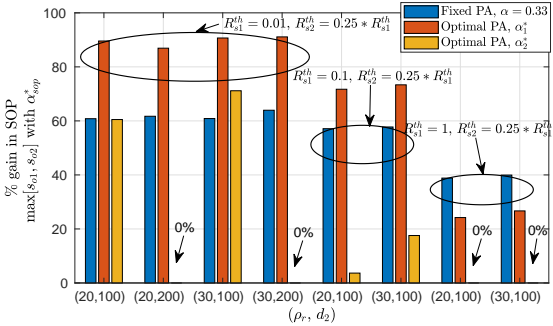}
\caption{Performance comparison of global-optimal PA $\alpha_{sop}^{*}$ with fixed PA, and individual optimal PA $\alpha_{1}^{*}$ and $\alpha_{2}^{*}$.}
\label{performance_comp}
\end{figure}
To analyze the performance gain obtained by the proposed protocol for maximizing secrecy fairness, Fig. \ref{performance_comp} demonstrates the performance comparison of globally optimized PA $\alpha_{sop}^{*}$ with fixed PA $\alpha=0.33$, and individual users' optimal PAs $\alpha_{1}^{*}$ and $\alpha_{2}^{*}$, respectively, obtained by minimizing $s_{o1}$ and $s_{o2}$. Results indicates the percentage gain which depicts that $\alpha_{sop}^{*}$ obtains best SOP performance, because of ensuring secrecy fairness between users. Note that the average percentage improvement by $\alpha_{sop}^{*}$ over fixed PA, optimal PAs $\alpha_{1}^{*}$ and $\alpha_{2}^{*}$ are approaximately $55.12\%$, $69.30\%$ and $19.11\%$, respectively.
\section{Concluding Remarks}
This paper has proposed a novel decoding order for a NOMA system with two untrusted users, that can provide positive secrecy rate to both users. With the objective of secrecy fairness between users, globally-optimized PA to minimize the maximum SOP between users is presented. Asymptotic solution is also obtained to gain analytical insights. Also, individual PAs minimizing SOPs for both the users, along with closed-form asymptotic expressions are presented. Numerical results are conducted to verify the correctness of analytical expressions as well as to provide insights on optimal performance and significant performance gains.

\bibliographystyle{IEEEtran}
\bibliography{ref}

\begin{thebibliography}{10}
\providecommand{\url}[1]{#1}
\csname url@rmstyle\endcsname
\providecommand{\newblock}{\relax}
\providecommand{\bibinfo}[2]{#2}
\providecommand\BIBentrySTDinterwordspacing{\spaceskip=0pt\relax}
\providecommand\BIBentryALTinterwordstretchfactor{4}
\providecommand\BIBentryALTinterwordspacing{\spaceskip=\fontdimen2\font plus
\BIBentryALTinterwordstretchfactor\fontdimen3\font minus
  \fontdimen4\font\relax}
\providecommand\BIBforeignlanguage[2]{{%
\expandafter\ifx\csname l@#1\endcsname\relax
\typeout{** WARNING: IEEEtran.bst: No hyphenation pattern has been}%
\typeout{** loaded for the language `#1'. Using the pattern for}%
\typeout{** the default language instead.}%
\else
\language=\csname l@#1\endcsname
\fi
#2}}

\bibitem{nomasurvey}
L.~{Dai}, B.~{Wang}, Z.~{Ding}, Z.~{Wang}, S.~{Chen}, and L.~{Hanzo}, ``A
  survey of non-orthogonal multiple access for 5{G},'' \emph{IEEE Commun.
  Surveys Tuts.}, vol.~20, no.~3, pp. 2294--2323, thirdquarter 2018.

\bibitem{liu2017enhancing}
Y.~Liu, Z.~Qin, M.~Elkashlan, Y.~Gao, and L.~Hanzo, ``Enhancing the physical
  layer security of non-orthogonal multiple access in large-scale networks,''
  \emph{IEEE Trans. Wireless Commun.}, vol.~16, no.~3, pp. 1656--1672, Mar.
  2017.

\bibitem{pls}
Y.~Wu, A.~Khisti, C.~Xiao, G.~Caire, K.~Wong, and X.~Gao, ``A survey of
  physical layer security techniques for 5{G} wireless networks and challenges
  ahead,'' \emph{IEEE J. Sel. Areas Commun.}, vol.~36, no.~4, pp. 679--695,
  Apr. 2018.

\bibitem{zhang2016secrecy}
Y.~Zhang, H.-M. Wang, Q.~Yang, and Z.~Ding, ``Secrecy sum rate maximization in
  non-orthogonal multiple access,'' \emph{IEEE Commun. Lett.}, vol.~20, no.~5,
  pp. 930--933, May 2016.

\bibitem{ding2017spectral}
Z.~Ding, Z.~Zhao, M.~Peng, and H.~V. Poor, ``On the spectral efficiency and
  security enhancements of {NOMA} assisted multicast-unicast streaming,''
  \emph{IEEE Trans. Commun.}, vol.~65, no.~7, pp. 3151--3163, July 2017.

\bibitem{8245831}
J.~{Chen}, L.~{Yang}, and M.~{Alouini}, ``Physical layer security for
  cooperative {NOMA} systems,'' \emph{IEEE Trans. Veh. Technol.}, vol.~67,
  no.~5, pp. 4645--4649, May 2018.

\bibitem{lv2018secure}
L.~Lv, Z.~Ding, Q.~Ni, and J.~Chen, ``Secure {MISO-NOMA} transmission with
  artificial noise,'' \emph{IEEE Trans. Veh. Technol.}, vol.~67, no.~7, pp.
  6700--6705, July 2018.

\bibitem{basepaper}
B.~M. {ElHalawany} and K.~{Wu}, ``Physical-layer security of {NOMA} systems
  under untrusted users,'' in \emph{Proc. IEEE GLOBECOM}, United Arab Emirates,
  Dec. 2018, pp. 1--6.

\bibitem{saini2016ofdma}
R.~Saini, D.~Mishra, and S.~De, ``{OFDMA}-based {DF} secure cooperative
  communication with untrusted users,'' \emph{IEEE Commun. Lett.}, vol.~20,
  no.~4, pp. 716--719, Apr. 2016.

\bibitem{saini2019subcarrier}
R.~{Saini}, D.~{Mishra}, and S.~{De}, ``Subcarrier pairing as channel gain
  tailoring: Joint resource allocation for relay-assisted secure {OFDMA} with
  untrusted users,'' \emph{Physical Communication}, vol.~32, pp. 217--230,
  2019.

\bibitem{7987792}
R.~Saini, D.~Mishra, and S.~De, ``Utility regions for {DF} relay in
  {OFDMA}-based secure communication with untrusted users,'' \emph{IEEE Commun.
  Lett.}, vol.~21, no.~11, pp. 2512--2515, Nov. 2017.

\bibitem{gc2019noma}
S.~Thapar, D.~Mishra, and R.~Saini, ``Novel outage-aware {NOMA} protocol for
  secrecy fairness maximization among untrusted users,'' \emph{submitted to
  IEEE journal}, Aug. 2019.

\bibitem{tse2005fundamentals}
D.~Tse and P.~Viswanath, \emph{Fundamentals of wireless communication}.\hskip
  1em plus 0.5em minus 0.4em\relax Cambridge university press, 2005.

\bibitem{boyd2004convex}
S.~Boyd and L.~Vandenberghe, \emph{Convex optimization}.\hskip 1em plus 0.5em
  minus 0.4em\relax Cambridge university press, 2004.

\bibitem{mishra2016joint}
D.~Mishra, S.~De, and C.-F. Chiasserini, ``Joint optimization schemes for
  cooperative wireless information and power transfer over rician channels,''
  \emph{IEEE Trans. Commun.}, vol.~64, no.~2, pp. 554--571, Feb. 2016.

\bibitem{bazaara1979nonlinear}
M.~Bazaara, H.~Sherali, and C.~Shetty, \emph{Nonlinear programming: theory and
  applications}.\hskip 1em plus 0.5em minus 0.4em\relax New York: Wiley, 1979.

\bibitem{chang2009n}
Y.-C. Chang, ``{N}-dimension golden section search: Its variants and
  limitations,'' in \emph{Proc. 2nd Int. Conf. on Biomedical Engineering and
  Informatics (BMEI)}, China, Oct. 2009, pp. 1--6.

\bibitem{ravindran2006engineering}
A.~Ravindran, G.~V. Reklaitis, and K.~M. Ragsdell, \emph{Engineering
  optimization: methods and applications}.\hskip 1em plus 0.5em minus
  0.4em\relax John Wiley \& Sons, 2006.

\end{thebibliography}
\end{document}